\documentclass[10pt,a4paper]{article}
\usepackage{jheppub}
\usepackage{amsmath}
\usepackage{graphicx}
\usepackage{amssymb}
\usepackage{amsfonts}
\usepackage{subfigure}
\usepackage{latexsym,epsfig}
\usepackage{slashed,xcolor}

\def\beq{\begin{eqnarray}}
\def\eeq{\end{eqnarray}}

\def\al{\alpha}
\def\be{\beta}

\def\ga{\gamma}

\def\la{\lambda}
\def\na{\nabla}
\def\pa{\partial}

\def\ph{\varphi}

\newcommand{\MET}{\textbf{\textit{E}$_{\rm T}^{\rm miss}$}}

\title{Torsion as a Dark Matter Candidate from the Higgs Portal}

\author[a,b,1]{Alexander S. Belyaev\note{e-mail: a.belyaev@soton.ac.uk}}
\author[c,d,2]{Ilya L. Shapiro\note{e-mail: shapiro@fisica.ufjf.br}}
\author[a,b,3]{Marc C. Thomas\note{e-mail: m.c.thomas@soton.ac.uk}}

\affiliation{School of Physics \& Astronomy, University of S., Southampton SO17 1BJ, United Kingdom}
\affiliation{Rutherford Appleton Laboratory,Science \& Technology Facilities Council (STFC),Chilton, Didcot. Oxon OX11 0QX, United Kingdom}
\affiliation{Departamento de F\'{\i}sica -- ICE,
Universidade Federal de Juiz de Fora,  Juiz de Fora, 36036-330, MG,  Brazil}
\affiliation{Tomsk State Pedagogical University and
Tomsk State University, Tomsk, 634041, Russia}

\abstract{
Torsion is a metric-independent component of gravitation, which
may provide a more general geometry than the one taking place
within general relativity.  On the other hand torsion could lead to
interesting phenomenology in both particle physics and cosmology.
In the present work it is shown that a torsion field
interacting with the SM Higgs doublet and having a negligible
coupling to SM fermions is protected from decaying by a $Z_2$
symmetry, and therefore becomes a promising Dark Matter (DM) candidate. In
order to check the consistency of  this scenario we evaluate the DM
relic density and explore direct DM detection and collider constraints
on this model. It turns out that in the model when the Higgs boson
is only partly responsible for the generation of torsion mass, there
is a region of parameter space where torsion contributes 100\%
to the DM budget of the Universe. Furthermore, we show that the LHC
currently has a limited sensitivity to the torsion parameter space 
via mono-jet signature
and will be able to considerably improve its coverage
of the torsion parameter space with the projected high luminosity.}


\preprint{}

\makeindex

\begin{document}
\maketitle

\section{Introduction}

The existence of Dark Matter(DM) has been established  beyond
reasonable doubt at the cosmological scale, and is one of the main pieces of evidence
of physics beyond the Standard Model(SM). There are several kinds
of realistic extension of the SM which can explain  the origin of DM.
For instance, assuming some scenario of Grand Unification of electroweak
and strong interactions at the UV scale, one naturally arrives at the
concept of new weakly interacting particles  which decouple from the
baryonic sector in the Early Universe. This scenario serves as a
theoretical grounding for models with weakly interacting massive
particles -- WIMPs, which appear in the majority of the theoretically motivated
scenarios which predict a DM candidate, such as SUSY with R-parity
\cite{Goldberg:1983nd,Ellis:1983ew}, Universal Extra Dimensions \cite{Antoniadis:1990ew, Appelquist:2000nn, Servant:2002aq, Csaki:2003sh}, Little Higgs \cite{ArkaniHamed:2002qx, Cheng:2003ju, Cheng:2004yc, Low:2004xc, Hubisz:2004ft, Cheng:2005as, Hubisz:2005tx} or Technicolor~\cite{Nussinov:1985xr,Barr:1990ca,Gudnason:2006ug}.

At the same time, one can consider different kinds of extensions of the SM
of gravitational origin. One example, which is well-explored, is
related to space-time torsion \cite{Hehl:1976kj,Shapiro:2001rz}.
An advantage of this kind of extension is that it follows from the
unification of matter fields with gravity, including in the framework
of (super)string theory. Indeed, superstring theory predicts the
existence of torsion with a non-minimal coupling to scalar fields
and fermions. As a result, the compactification of extra dimensions
can in general give rise to a theory with a modified form of such
non-minimal couplings in the low-energy limit.

At the GUT scale one has to use the field theory description instead of
string theory, and therefore the first problem  is to produce a
consistent effective description of torsion and its interaction
with matter fields. The last issue was addressed in the papers
\cite{Buchbinder:1985ux,Buchbinder:1990ku} (see also
\cite{Buchbinder:1992rb,Shapiro:2001rz} for the case when
torsion is a classical background for quantum matter fields).
The main theoretical result was that the consistent formulation
of such a theory requires a non-minimal interaction of the axial-vector
component of torsion with fermions and also with scalar
constituents of the theory. The non-minimal interactions are
characterised by several free parameters, which can be defined
only from experiments or observations. As far as these new
parameters are introduced, the theory is consistently formulated
at the semi-classical level.

The most difficult problem is related to formulating the theory of
dynamical torsion which can provide a consistent effective quantum
field theory description for the interaction with fermions of the SM.
This issue was discussed in Refs.
\cite{Belyaev:1997zv,Belyaev:1998ax}, where it was shown that
the consistent theory of antisymmetric torsion is described by
the Proca-like action of axial vector fields. This results served as
the basis for phenomenological studies (see also earlier work
\cite{Carroll:1994dq}), including an exploration of the LHC potential
to probe torsion field \cite{Belyaev:2007fn,Belyaev:1997zv,Belyaev:1998ax}.

In the present work we go beyond this framework and explore the
phenomenological consequences of the non-minimal interaction
between torsion and scalar fields. In particular we stress that in
the case when the Higgs-fermion couplings vanish and torsion interacts
with the SM particles only through the Higgs fields, torsion is protected
from decaying by a $Z_2$ symmetry. Therefore in this case torsion
becomes a good DM candidate. Torsion can also play the role of DM
when its  couplings to fermions are highly suppressed by the Planck
mass, such that torsion's life time is of the order of the age of the
Universe. In this paper we explore the potential of DM direct
detection experiments as well as the LHC to probe the parameter
space of this scenario.

The paper is organised as follows. In Sect. 2 we present a brief
survey of the theory  related to torsion interactions with
fermions and scalars. Furthermore, a mechanism of generating 
relatively large non-minimal parameters for the torsion-Higgs 
interaction is sketched. In Sect. 3 we describe a torsion-based 
model of DM and in Sect. 4 discuss the possible potentials of the 
LHC, DM direct detection and Planck experiments to probe
the scenario with torsion being a DM candidate. Finally, in
Sect. 5 we draw our conclusions.

\section{Interactions of torsion with fermions and  scalars}
\label{Sect2}

Torsion $\,T^\al_{\;\;\be\ga}\,$ is a metric-independent 
tensor field which is defined as
$$
{\Gamma}^\alpha_{\;\beta\gamma} -
{\Gamma}^\alpha_{\;\gamma\beta} =
T^\alpha_{\;\;\beta\gamma}\,.
$$
One can split torsion into irreducible components
\beq
T_{\alpha\beta\mu} =
\frac{1}{3} \left( T_{\beta}g_{\alpha\mu} -
T_{\mu}g_{\alpha\beta} \right)
- \frac{1}{6} \varepsilon_{\alpha\beta\mu\nu}
\,S^{\nu} + q_{\alpha\beta\mu}\,,
\label{t1}
\eeq
where the axial vector \ $S^\nu$ \ is dual to the completely
antisymmetric part of the torsion tensor
$\;S^{\nu} = \epsilon^{\al\be\mu\nu}T_{\al\be\mu}$,
$\,T_\al=T^\la_{\,\,\,\al\la}\,$ is a
vector trace of torsion and \ $q_{\alpha\beta\mu}$ \ is a
tensor component.

The most general non-minimal action for a Dirac spinor coupled to
torsion is
\beq
S_f \,=\, \int\sqrt{-g}\,\left\{\,i\bar{\psi}\gamma^\mu
\big( \na_\mu - i \eta_1\gamma^5S_\mu + i\eta_2T_\mu\big)\psi
- m\bar{\psi}\psi \right\}\,,
\label{t2}
\eeq
where $\eta_1,\,\eta_2$ are non-minimal parameters and $\na_\mu$
is Riemannian covariant derivative, constructed without torsion. The
minimal interaction corresponds to the values $\,\eta_1 = -1/8$ and
$\eta_2 = 0$.  However, in the theory which includes scalar fields
coupled to fermions via Yukawa interactions, the minimal theory is
not renormalisable \cite{Buchbinder:1985ux} and hence one has to
introduce the non-minimal coupling of both fermions and scalars
with the background torsion field. For instance, this can be seen
from the renormalisation group equation for $\eta_1$,
\beq
\mu\,\frac{d\eta_{1,2}}{d\mu}\,=\,
\big( C_1\,h^2 \,+\, \mbox{higher loop contributions}\big)
\cdot\eta_{1,2}\,,
\label{RG for eta}
\eeq
where $h$ is Yukawa coupling and the coefficient $C_1$ is
model-dependent. It is clear that the minimal value $\eta_1=1/8$
is not stable under quantum corrections, hence one has to assume
an arbitrary $\eta_1$. At the same time, $\eta_2=0$ does not lead
to such a problem, therefore for the sake of simplicity one can
restrict consideration by the purely antisymmetric torsion and
the unique parameter $\eta_1$.

In the scalar sector one meets the following non-minimal action:
\beq
S_0=\int d^4\sqrt{-g}\,\Big\{\,
\na^\mu\ph^*\,\na_\mu\ph
- m^2\left|\ph\right|^2
+ \sum_{i=1}^{5}\xi_i\,P_i\,\left|\ph\right|^2
- f\left|\ph\right|^4\Big\}\,,
\label{scalar1-nm}
\eeq
where \cite{Buchbinder:1985ux,Buchbinder:1990ku}
\beq
P_1 = R, \,\,\,\,\,\,
P_2 = \na_\al\,T^\al, \,\,\,\,\,\,
P_3 = T_\al\,T^\al, \,\,\,\,\,\,
P_4 = S_\al\,S^\al, \,\,\,\,\,\,
P_5 = q_{\al\be\ga}\,q^{\al\be\ga}\,.
\label{Pi}
\eeq
In general, there can be up to five non-minimal parameters
$\xi_{1...5}$. However, if we restrict our attention to the
antisymmetric torsion and flat space-time, the only one
relevant parameter is $\xi_4$. The analysis of one-loop
corrections shows that the non-minimal parameter $\xi_4$
and $\eta_1$ in the fermion sector are closely related, for
instance the renormalisation group equation for $\xi_4$ has
the general form
\beq
\mu\,\frac{d\xi_4}{d\mu}\,=\,C_2\,h^2\,\eta^2_1
\,+\,\big(C_3 g^2 + C_4 g^2 + C_5 f\big)\xi_4
\,+\,
\mbox{higher loop contributions}
\,,
\label{RG for xi}
\eeq
with $\,g,h,f\,$ being gauge, Yukawa and scalar coupling constants
(in the more realistic case, such as the SM, there will be several gauge
and Yukawa-dependent terms) and model-dependent coefficients
$C_{2,..,5}$ \cite{Buchbinder:1985ux,Buchbinder:1992rb}. Let us note that all that we
have discussed until now corresponds to the case of a background torsion,
which does not depend on the action of torsion. However, complete
consideration should take into account the dynamics of torsion itself.

One can establish the action for the propagating torsion by
requesting consistency of the effective low-energy quantum theory
of torsion coupled to fermions and scalars, e.g., of the SM.
Unitarity of the low-energy theory leads to a possible form
of the torsion action
\cite{Belyaev:1997zv,Belyaev:1998ax}
\beq
{\cal S}_{tor}^{TS-kin}\,=\,\int d^4 x
\,\Big\{\,-\frac14\,S_{\mu\nu}S^{\mu\nu}
+ \frac12\,M_{ts}^2\, S_\mu S^\mu\,\Big\}\,,
\label{action}
\eeq
where 
$\,S_{\mu\nu}=\pa_\mu S_\nu-\pa_\nu S_\mu$ and $M_{ts}$ is
a torsion mass. Furthermore, one can show that quantum corrections
preserve unitarity only if the relation
\beq
\frac{M^2_{ts}}{\eta_1} &\gg&  m_f^2
\label{crit}
\eeq
is satisfied for all fermion fields with masses $m_f$ \cite{Belyaev:1997zv,Belyaev:1998ax,deBerredoPeixoto:1999vj}. In Ref.
\cite{Chang:2000yw,Lebedev:2002dp} similar constraints in the 
theory with extra dimensions were discussed.

The consistency relations (\ref{crit}) concern torsion mass and
non-minimal parameters of the fermion-torsion interaction. This
condition can be satisfied either for a huge torsion mass or for a
very weak fermion-torsion interaction. At this point we can conclude
that the only torsion-related parameter which is not restricted
is $\xi_4$ in Eq. (\ref{scalar1-nm}). From the theoretical side there
are no reasons to impose such a restriction. At the
same time, there are no phenomenological constraints on $\xi_4$,
because the  present-day magnitude of the torsion field is extremely small
\cite{Kostelecky:2010ze}. Therefore it would be interesting to explore
the physical consequences of different values of $\xi_4$, including
of relatively large (compared to $\eta_1$) magnitudes, because they 
are not ruled out by the condition of theoretical consistency and, as 
we shall see below, may be phenomenologically fruitful.

Before we go to the phenomenological part in the next sections,
let us present some additional theoretical arguments. One has
to remember that $\xi_4$ is a free parameter and its value can
be determined only by comparison to experimental or observational
data. However, it is possible to sketch a situation when a relatively
large value of $\xi_4$ may emerge\footnote{A Riemannian version of a
similar arguments has been discussed recently in \cite{Netto:2015cba}
in relation to the QFT-based mechanism for inflation.}.

In the usual GUT models the SM Higgs couples not only to SM
fermions, but also to other fields, which belong to the extension of
the $SU(3)_c\times SU(2)\times U(1)$ gauge group. These fermions
are supposed to have much greater masses compared to the SM
particles. At the same time there is another possibility to decouple
GUT particles from
those of the SM. Namely, one can assume that the hidden UV sector
of the GUT model is characterised by strong interactions, and that
at an energy scale which is much lower than the GUT scale, these
fields get confined
into composite particles which have quantum numbers distinct from the
original elementary particles. As a result they decouple from the SM
particles. After such a decoupling only gravitational interactions are possible,
hence these composite particles can constitute some part of the DM. In
what follows we discuss the situation when the main part of the DM is
composed by torsion itself. An advantage of this scheme is that torsion
is a geometric field and, hence, may have very weak non-gravitational
interaction to the SM particles.

An important aspect of the IR-confined GUT is that close to the
decoupling scale there are typically very strong interactions and the
quantum fields system is in the strongly non-perturbative regime.
Then the specific GUT gauge, Yukawa and scalar couplings 
may be strong and one can observe an intensive running
in Eq. (\ref{RG for eta}) for the hidden fermions.

In this scenario the values of $\eta_1$ for the hidden fermions
may greatly increase due to the intensive running. Since the
masses of these fermions are supposed to be huge and,
moreover, they are eventually confined into hidden composite
particles, a possible violation of the criterion (\ref{crit})
does not put into danger the unitarity of the theory. But, as
a consequence of interacting with the SM Higgs, Eq.
(\ref{RG for xi}) may produce a short-period but very intensive
growth of the value of $\xi_4$ on the running from UV to IR.
Therefore, within this scheme there is nothing wrong with assuming
that at the low-energy of $\xi_4$ is ``unnaturally'' large. At the
same time this process does not concern at all the SM fermions,
hence the corresponding parameters $\eta_1$ remain very small.
In the next sections we will discuss some phenomenological
advantages of these assumptions.

\section{Torsion as a DM candidate}
\label{Sect3}

Starting from this section, we shall use simplified notations
$\xi_4=\xi$ and  $\eta_1=\eta$. According to the previous
considerations, the torsion-matter interaction has the form
\beq
{\cal L}_{tor}^{matter}
&=&
\eta\sum_i\bar{\psi_i}\gamma^\mu\gamma^5\psi_i S_\mu\,,
\label{action-matter}
\eeq
where $\psi_i=e,\mu,\tau,\nu_e,\nu_\mu,\nu_\tau,u,c,t,d,s,b$

Furthermore, the Higgs portal looks as follows:
\beq
{\cal L}_{tor}^{H}
&=&
\xi H^2 S^\mu S_\mu\,.
\label{action-H}
\eeq

One can see that the Higgs mechanism generates the mass
$M_{ts}^2=\xi v^2$ for the torsion $S^\mu$. The simplest scenario,
in which this mechanism is the only source of the $S^\mu$-field mass,
will be abbreviated hereafter as $TDM1$,  since in such a scenario
the torsion mass, $M_{ts}$, is a function of only one parameter
$\xi$. At the same time torsion can receive an additional mass contribution,
which we denote as $M'_{ts}$ which may come from the assumed
symmetry breaking at the high energy scale. Therefore, one can write
the general mass relation in the form
\beq
M_{ts}^2 &=& \xi v^2+{M'}_{ts}^2\,.
\eeq
We will abbreviate this (more general) scenario as $TDM2$
since $M_{ts}$ in this case depends on two parameters.
We have implemented both the $TDM1$ and $TDM2$ models into the CalcHEP package~\cite{CALCHEP}.

\begin{figure}[htb]
\centering
\includegraphics[width=0.9\textwidth]{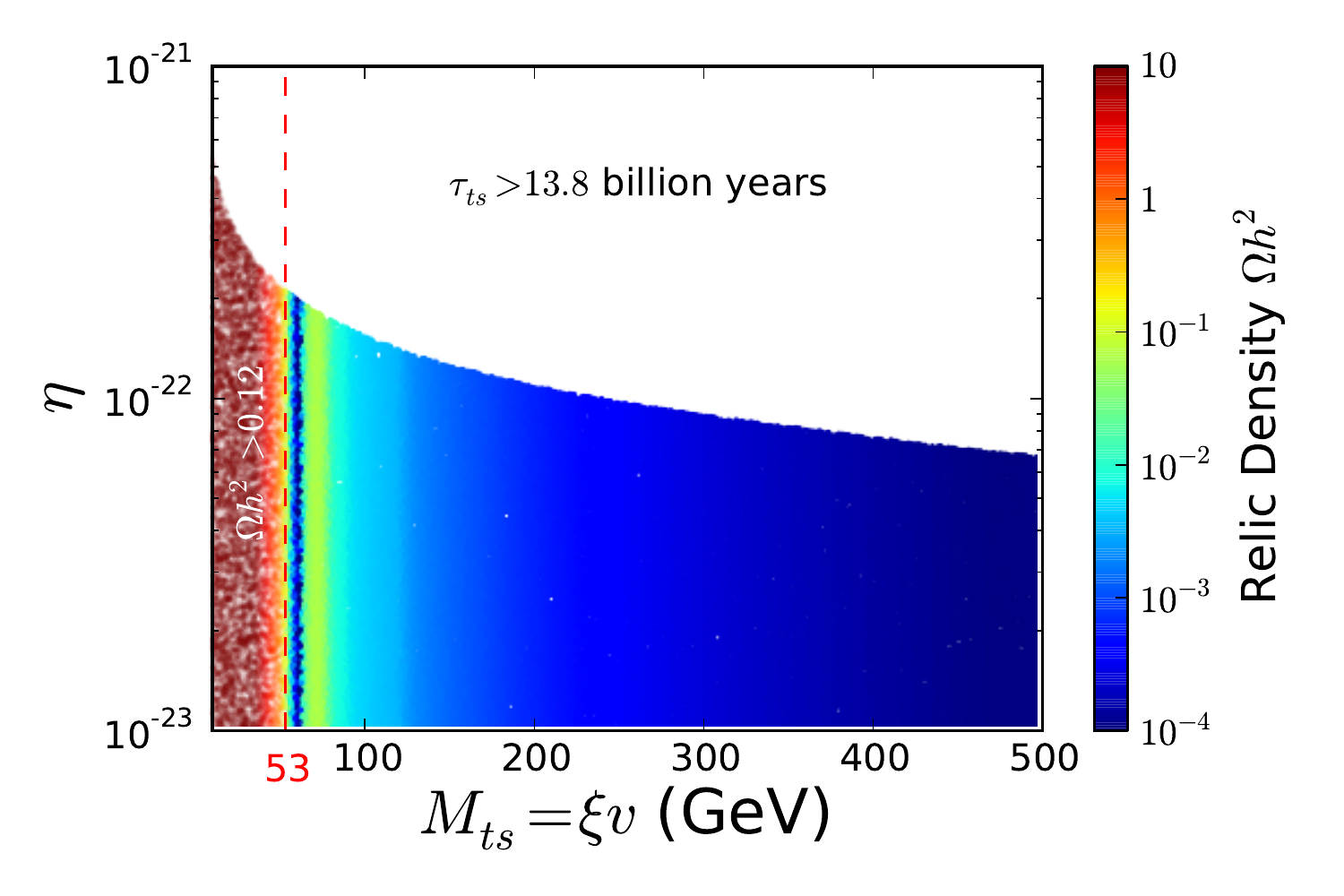}
\caption{
The upper limit on $\eta$ shown in the ($\eta-M_{ts}$) plane. 
The coloured region is allowed and presents the torsion 
relic density in standard units of $\Omega h^2$. The white region 
above the coloured one is excluded because $\tau_{ts}<13.8$~billion 
years.}
\label{fig:eta}
\end{figure}

Because we are exploring  the scenario where torsion is  stable at
the cosmological scale, its lifetime should be above 14 billion years, which
translates into a torsion width below $10^{-42}$ GeV.
This takes place for very small values of $\eta$, the non-minimal
coupling of $S^\mu$ to fermions, the allowed region for which is
indicated by coloured region in Fig~\ref{fig:eta}
in the ($\eta-M_{ts}$) plane. The white region is excluded
 because in these scenarios
$\tau_{ts}<13.8$~billion years.
 One can see that
this condition excludes values of $\eta>10^{-21}$. Hereafter we will be
mainly interested in the torsion-scalar coupling. Therefore, since $\eta$
is so small, one can assume that it is zero, or that $\eta$ is below the
aforementioned limit.
%
%

In this figure we also present a colour map of the torsion-DM relic density in standard units of $\Omega h^2$
which we evaluated using {\texttt{micrOMEGAs 4.2.5}}~\cite{Belanger:2013oya,Belanger:2006is, Belanger:2010gh}.
Precise measurements of the DM relic density from
Planck~\cite{Ade:2013zuv,Planck:2015xua} (and previously from WMAP~\cite{Hinshaw:2012aka}),
\begin{equation}
\Omega_{\rm DM}^{\rm Planck} h^2=0.1184\pm0.0012,
\label{eq:planck-limit}
\end{equation}
provide further constraints on the torsion parameter space.
Taking this additional measurement into account, we have found
that $M_{ts}$ below 53 GeV is excluded in the $TDM1$ scenario.
This is  related to the fact that for such low $M_{ts}$ and respectively 
low values of $\xi=M_{ts}/v^2$, 
torsion DM in the early Universe
annihilates into SM particles through an s-channel off-shell
Higgs boson exchange. For $M_{ts}$ below 53 GeV, the suppression of this DM annihilation cross section
leads to an overabundance of DM which is excluded by Planck data.
The excluded parameter space is located to the left side of the dashed red line.

At the same time, once the mass of the torsion is close to $M_H/2$,
the effective annihilation through the on-shell Higgs boson resonance
$S^\mu S^\mu \to H $ takes place. Since the cross section of this process
is high, the relic density is about 2-3 orders of
magnitude below the $\Omega_{\rm DM}^{\rm Planck} h^2$. This
region of parameter space, as  well as  other  regions with low
relic density are not necessarily excluded, since there could be additional
(non-torsion) sources of DM
contributing to the overall DM relic density.

We further constrain the torsion parameter space by using
LHC bounds on the invisible Higgs boson decay,
\begin{equation}
Br(H\to invisible) = Br(H\to S^\mu S^\mu) < 28\%
\end{equation}
at the 95$\%$ confidence level (CL)~\cite{Aad:2015txa}.
In addition,
we also check the spin-independent (SI)  cross section of DM
scattering off the nuclei relevant to DM direct detection (DD)
searches and the respective current exclusion by the
LUX~\cite{Akerib:2013tjd} collaboration. We express the LUX
sensitivity using the measure,
\beq
R_{SI}
&=&
\frac{\tilde\sigma_{SI}}{\sigma_{SI}^{LUX}}
\,=\,
\frac{\sigma_{SI}}{\sigma_{SI}^{LUX}}
\,\,\frac{\Omega h^2}{\Omega_{\rm DM}^{\rm Planck} h^2}
\eeq
which contains the ratio of SI  cross section  for DM 
scattering on the nuclei and the LUX limit on the cross section 
multiplied by a DM relic density re-scaling factor
$\,\frac{\Omega h^2}{\Omega_{\rm DM}^{\rm Planck} h^2}$,
 to take into account the case when torsion relic density only partly
 contribute to the overall DM relic density of the Universe.
Regions of torsion parameter space will be excluded if $R_{SI}>1$.

\begin{figure}[htb]
\centering
\includegraphics[width=0.9\textwidth]{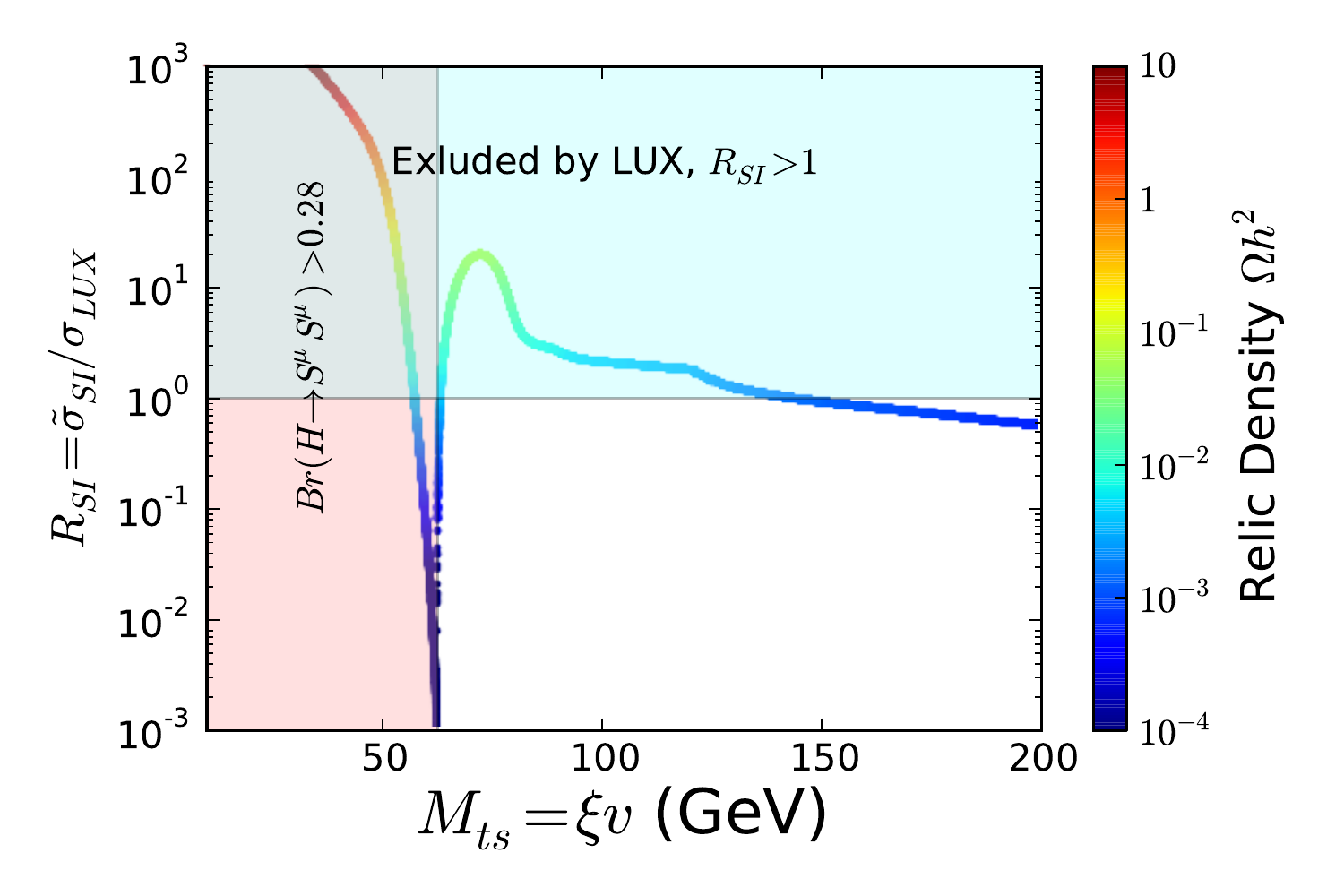}
\caption{Relic density, $Br(H\to S^\mu S^\mu)$ constraint and DM DD rates in the $TDM1$ scenario. The coloured line indicate
the value of DM DD rate relative to the LUX constraint, $R_{SI}$ which excludes
the model parameter space above  $R_{SI}=1$ level (green-shaded area).
At the same time the colour of this line indicate the value of the relic density,
$\Omega h^2$ which exceeds Planck constraint at the $M_{ts}\simeq 53$,
which therefore excludes parameter space below this value.
In the pink area, the torsion parameter space is excluded by 
constraints on $Br(H\to S^\mu S^\mu)$ from the Higgs invisible decay searches.
\label{fig:1dconstraints}}
\end{figure}

The effect of the Relic density, $Br(H\to S^\mu S^\mu)$
 and DM DD constraints are presented in Fig.~\ref{fig:1dconstraints}.
From this figure one can learn the following:
\\
1) The region $M_{ts}<M_H/2=62.5 {\rm\ GeV}$ is excluded by
the constarint $Br(H\to S^\mu S^\mu)$;
\\
2) The regions
$M_{ts}\lesssim 55{\rm\ GeV}$ and $M_H/2<M_{ts}\lesssim 145{\rm\  GeV}$
are excluded by LUX DD constraints;
\\
3) The region $M_{ts}\lesssim  53{\rm\ GeV}$ is excluded by the relic density
constraint, since in this region the torsion relic density is above the Planck
limit.

\begin{figure}[htb]
\subfigure[]{\includegraphics[width=0.6\textwidth]
{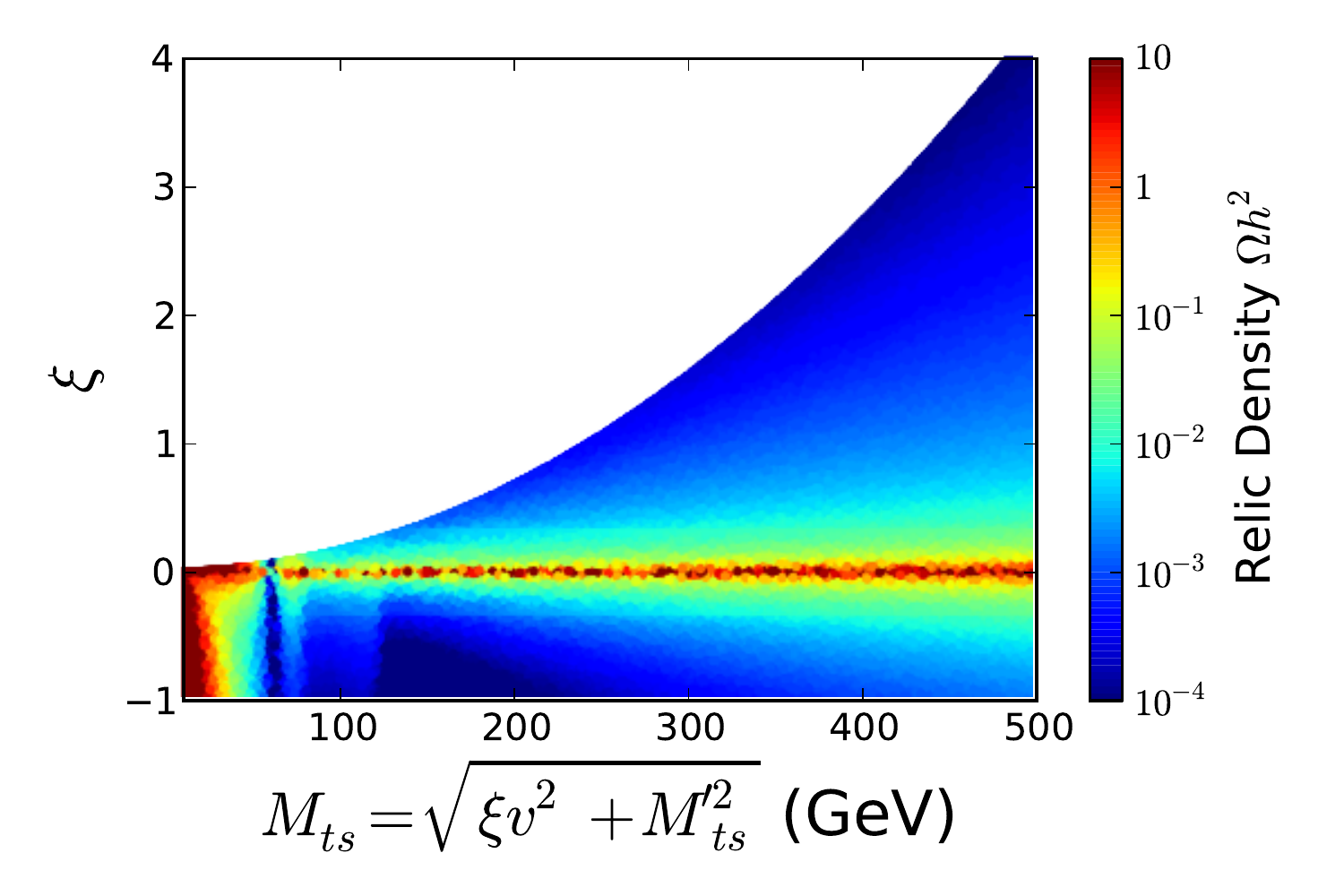}}%
\hspace{-1.8cm}
\subfigure[]{\includegraphics[width=0.6\textwidth]
{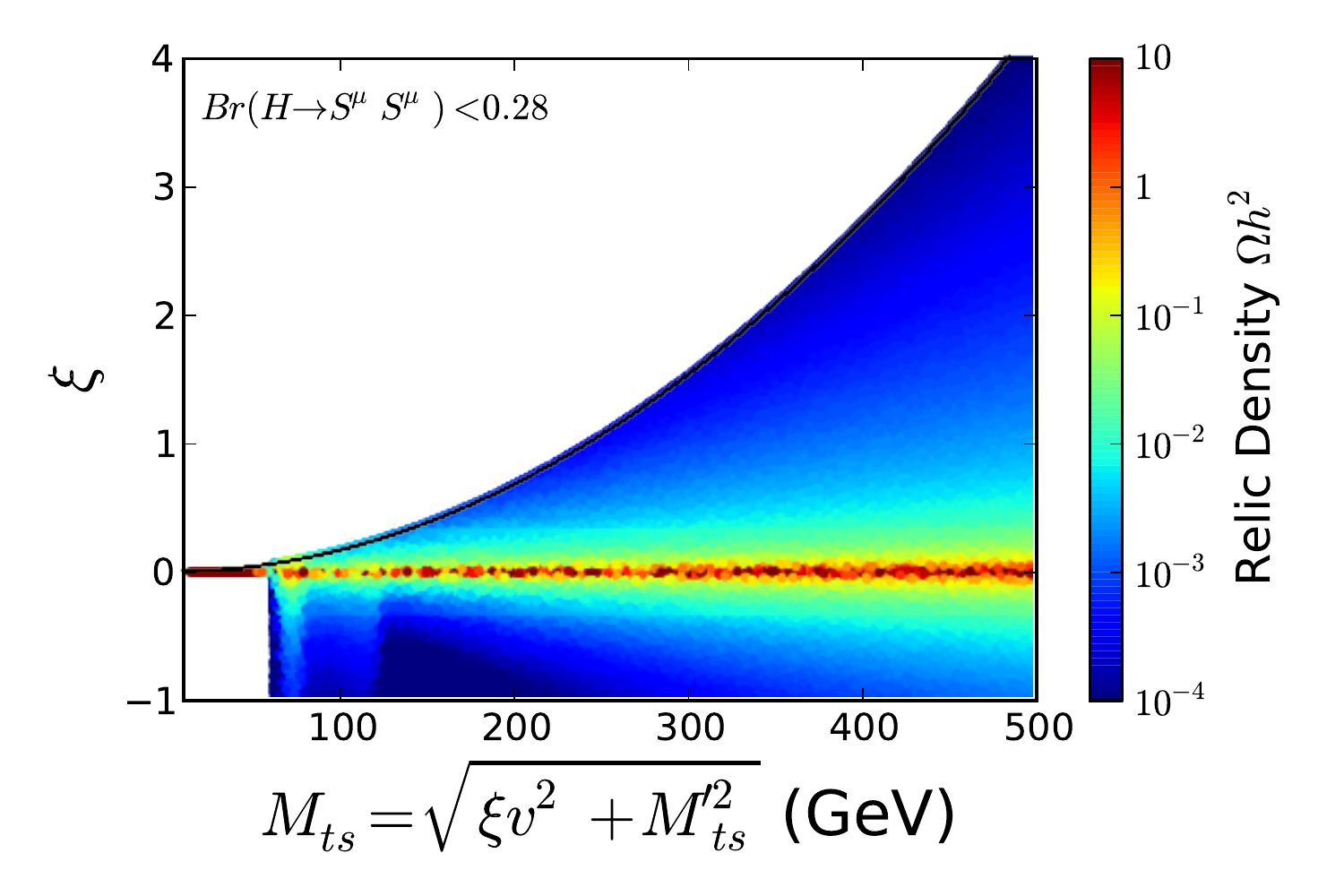}}
\\
\subfigure[]{\includegraphics[width=0.6\textwidth]
{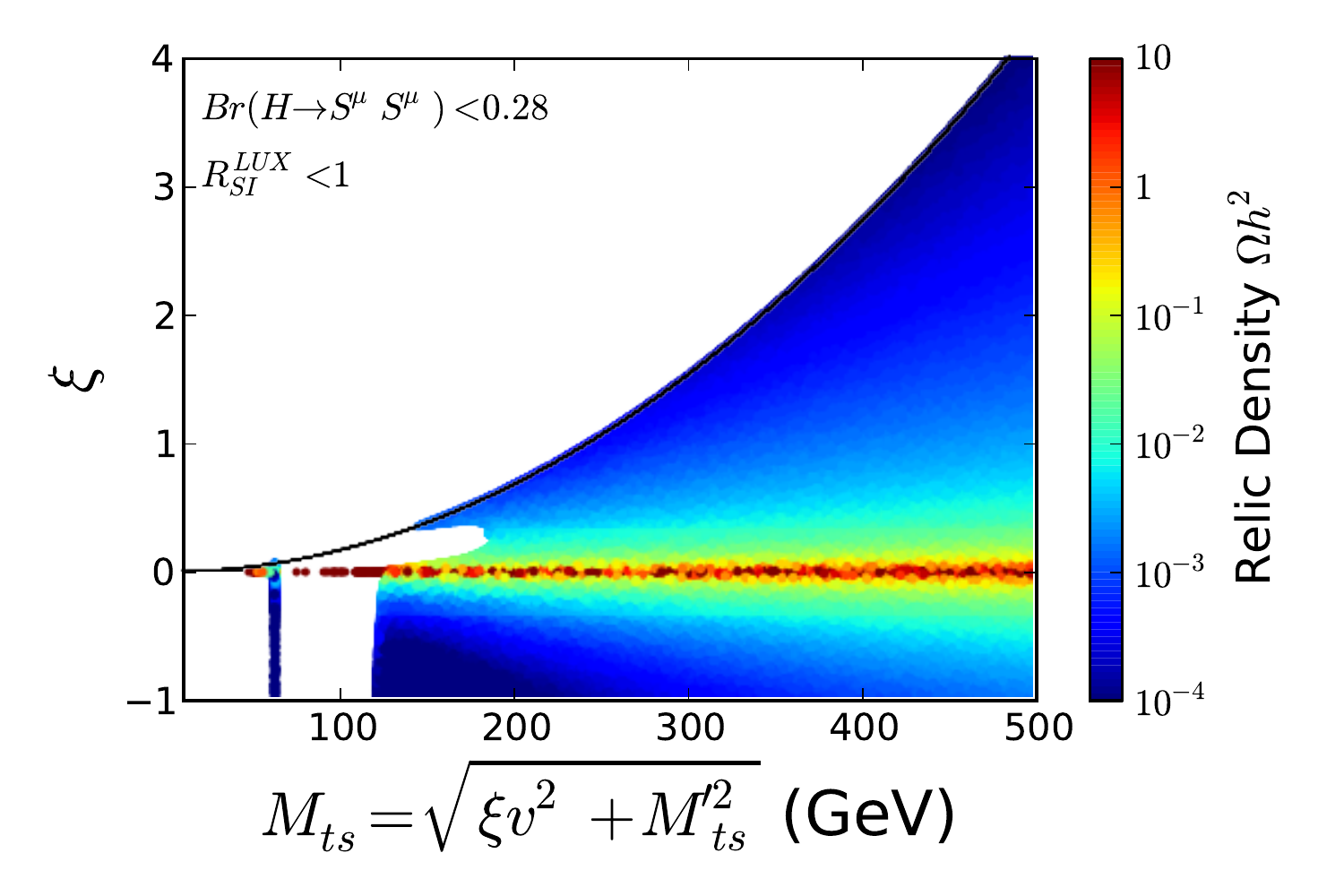}}%
\hspace{-1.8cm}
\subfigure[]{\includegraphics[width=0.6\textwidth]
{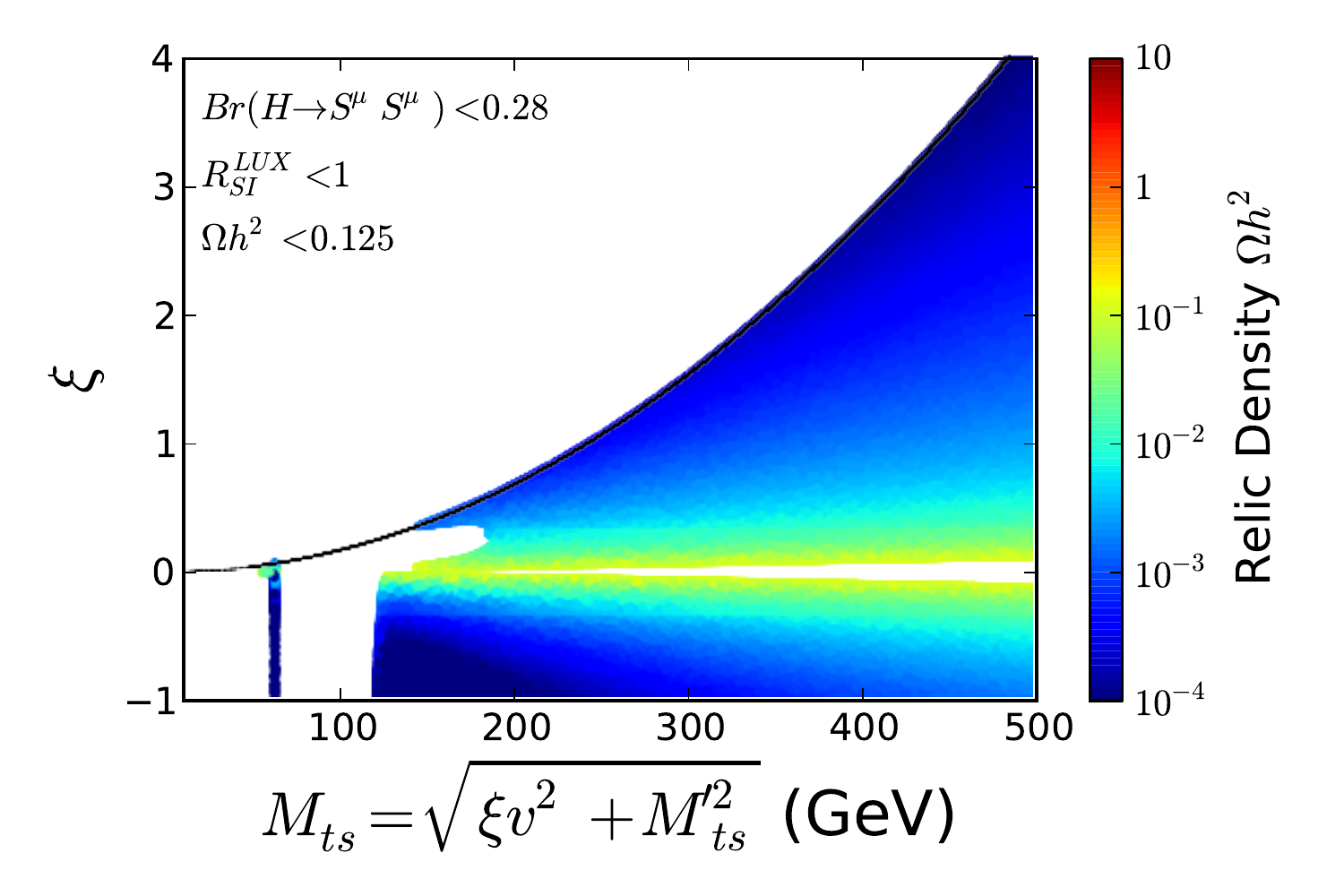}}%
\caption{The $\Omega h^2$ colour map in the
$\xi-M_{ts}$ parameter space of the $TDM2$ scenario
which survives the respective constraints:
a) the unconstrained parameter space; b)
$Br(H\to S^\mu S^\mu)<0.28$ constraint applied; c)
$R_{SI}$ constraint applied in addition; d)  upper  bound  on
$\Omega h^2$ applied on the top of the previous constraints.
The white colour indicates the excluded region, while the coloured area
is allowed by the set of constraints indicated for each frame.
\label{fig:2dconstraints}}
\end{figure}

Furthermore, the shape of the $R_{SI}$ line (which is correlated with the relic
density behaviour) exhibits several representative peaks and troughs, namely:
\\
a) the first sharp deep at $M_H/2$ corresponds to the resonant
DM annihilation through the Higgs boson which consequently leads to a sharp drop
in the relic density;
\\
b) the relic density sharply rises as soon as
its mass is above  $M_H/2$ because it no longer annihilates via an
exactly on-shell Higgs in the early Universe;
\\
c) once $M_{ts}$ approaches $M_W$ and $M_Z$, the
relic density drops again because the $S^\mu S^\mu\to W^+ W^-/ZZ$
annihilation processes become available;
\\
d) Finally, once $M_{ts}$ approaches $M_H$,
one can see another step down in relic density due to the
$S^\mu S^\mu\to HH$  processes.
As $M_{ts}$ increases further beyond $M_H$, the $\xi$ coupling
respectively increases leading to a further decrease
in the relic density of the $TDM1$-model.

The combination and complementary of these constraints ($Br(H\to S^\mu S^\mu)$, $\Omega h^2$
and DM DD) allow scenarios where $M_{ts}\gtrsim 145{\rm\ GeV}$, as well as
a very narrow region with $M_{ts}$ just
above  62.5 GeV (the Higgs resonance annihilation).
In both allowed regions the relic
density is one to two orders of magnitude below $\Omega_{\rm DM}^{\rm Planck} h^2$,
and therefore in the one-parametric $TDM1$
scenario torsion never contributes 100\% to the total budget of DM.

Let us now consider $TDM2$ scenario, corresponding to
more general case  when  torsion acquires an additional
mass, $M'_{ts}$, from symmetry breaking(s) at a higher energy scale, such
as in a GUT.  Then one can consider this mass as an additional parameter, which
would allow torsion to make up 100\% of the DM budget.
In Fig.~\ref{fig:2dconstraints}, we present
the $TDM2$ parameter space in the $\xi-M_{ts}$ plane,
following application of constraints from
$Br(H\to S^\mu S^\mu)$, $\Omega h^2$
and DM DD.
In this case
\beq
M_{ts}
&=&
\sqrt{\xi v^2 + {M'}_{ts}^2}\,,
\label{TDM2mass}
\eeq
and $\xi$ and $M_{ts}$ can be regarded as independent parameters.

\begin{figure}[htb]
\centering
\includegraphics[width=0.8\textwidth]{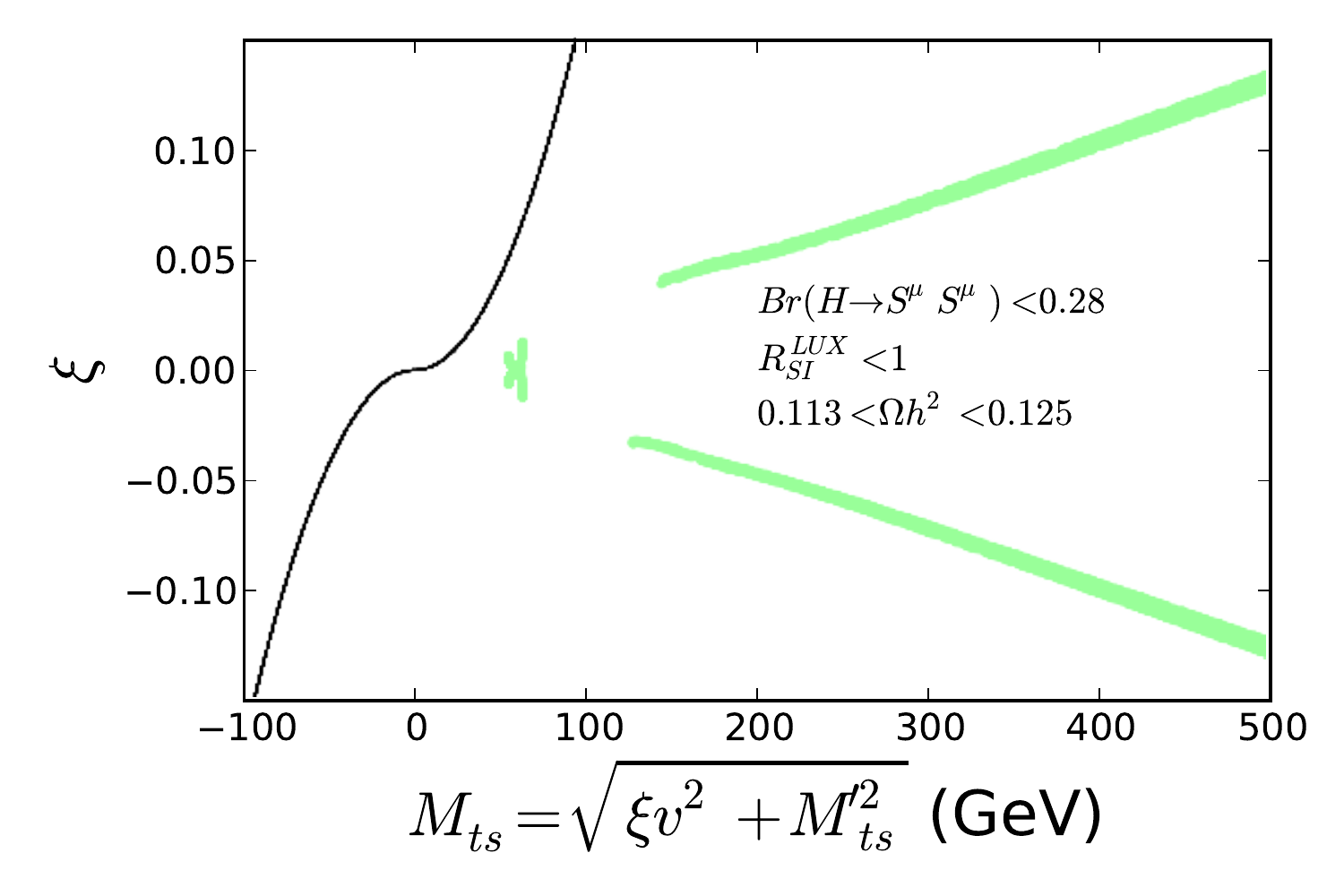}
\caption{\label{fig:2dfinal} The $TDM2$ parameter space which
survives
$Br(H\to S^\mu S^\mu)<0.28$, $R_{SI}<1$ and $0.113<\Omega h^2<0.125$ cuts. The allowed region is shown in green.}
\end{figure}

Fig.~\ref{fig:2dconstraints}(a) presents the colour map of $\Omega h^2$
in the unconstrained parameter space, Fig.~\ref{fig:2dconstraints}(b)
demonstrates the effect of the $Br(H\to S^\mu S^\mu)<0.28$ constraint and
Fig.~\ref{fig:2dconstraints}(c) shows the effect of the additional
application of  $R_{SI}$ constraint. Finally, Fig.~\ref{fig:2dconstraints}(d)
presents the parameter space after applying an upper  bound
on $\Omega h^2$.
The coloured parameter space indicates the parameter space which survives
the respective constraints. The upper edge of the coloured space indicated
by  the black contour corresponds to the $TDM1$ scenario.
One can see that the allowed  $M_{ts}$ mass range looks very similar  to
the $TDM1$ case.
However, contrary to the $TDM1$ scenario, for any mass $M_{ts}$
one can find in the $TDM2$ parameter space value(s) of the $\xi$ 
parameter for which torsion contributes 100\% to the DM relic 
density budget.
This $TDM2$ scenario, where all constraints are satisfied, including that
the entire relic density is explained with
torsion (including the {\it lower} cut on $\Omega h^2$ not present in
Fig.~\ref{fig:2dconstraints}(d)) is denoted by the green-coloured region
in Fig.~\ref{fig:2dfinal}.
One can see that even for $M_{ts}\simeq M_H/2$ one can have
$\Omega h^2 \simeq 0.112$, i.e., it is possible to find scenario
when torsion is the sole Dark Matter.
This scenario can be realised for small values of $\xi$ ($\simeq 0.01$)
and the values of the GUT contribution to torsion mass, ${M'}_{ts}$,
which are close to $M_H/2$. One can also see that when $M_{ts}\gtrsim 135$~GeV,
there are regions of parameter space, represented by the green bands symmetric about the $\xi=0$ line,
where torsion provides 100\% of the relic density budget.

\section{Large hadron collider sensitivity to the torsion DM}

\begin{figure}[htb]
{\includegraphics[width=\textwidth]{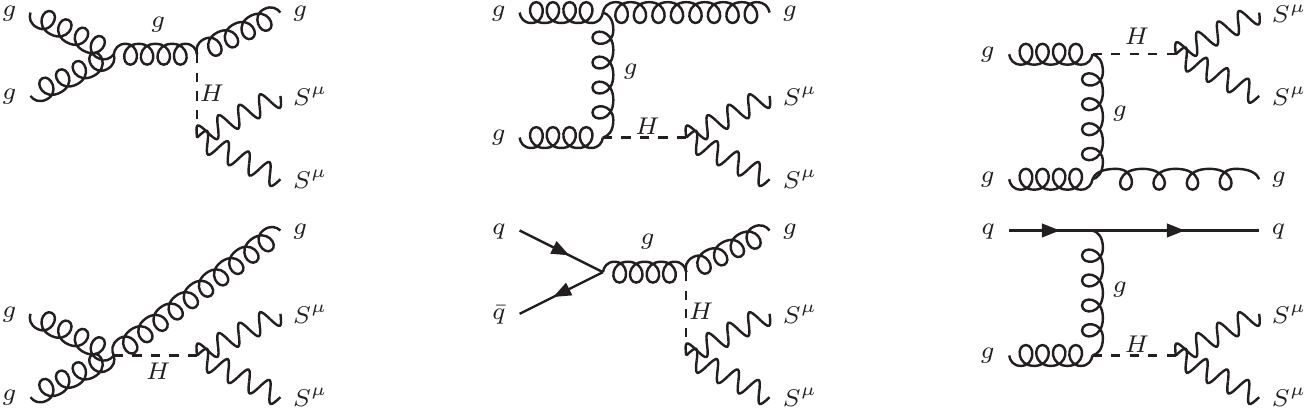}}
\caption{\label{fig:fd}
Feynman diagrams for the mono-jet signature $pp\to S^\mu S^\mu jet$ 
of the torsion pair production at the LHC.}
\end{figure}

In this section we find the current and project limits from the LHC on the torsion DM parameter space.
Torsion, being a DM, gives rise mono-jet signatures at the LHC, $pp\to S^\mu S^\mu jet$,
when a pair of torsion particles is recoiled against the hard 
quark or gluon jet coming from the initial state radiation.
Feynman diagrams for this process are presented in Fig.~\ref{fig:fd}.
The relevant parameter space for the model is eventually two dimensional,
consisting of the $\xi$ and $M_{ts}$ parameters. There is an upper limit on the
value of $\xi=\xi_{max+}=M_{ts}^2/vev^2$, coming from the $M_{ts}'>0$ requirement.
There is no limit on the absolute value of $\xi$ for $\xi<0$, except
the perturbativity one.
\begin{figure}[htb]
\centering
{\includegraphics[width=0.85\textwidth]{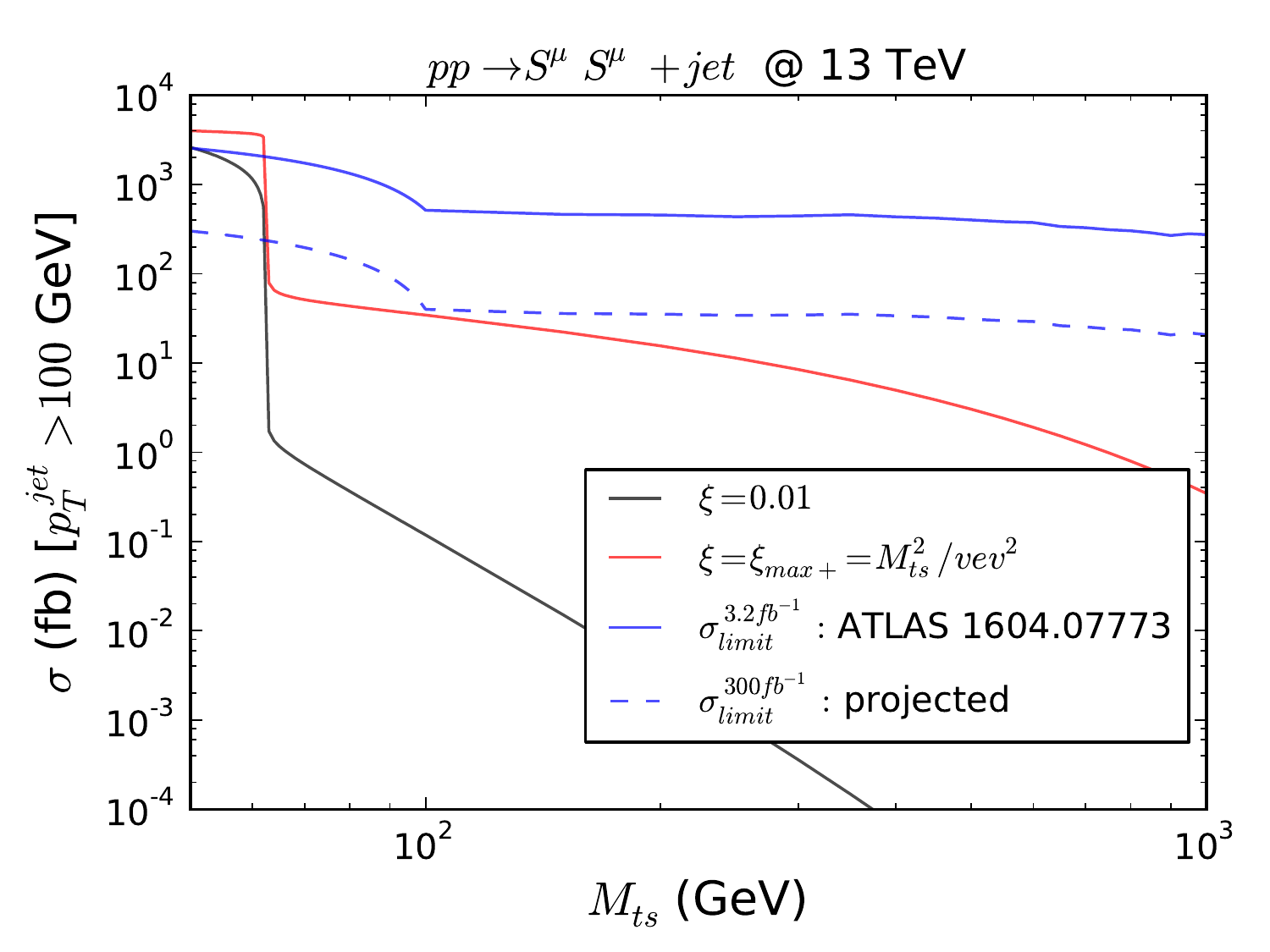}
}
\vskip -0.5cm
\caption{\label{fig:cs}Cross sections for the torsion  mono-jet
signature versus $M_{ts}$ at LHC@13 TeV with the cut $\MET>100$~GeV
for two values of $\xi$: $\xi=0.01$ (solid black) and
$\xi=\xi_{max+}$ (solid red)
as well as  LHC  limits obtained using an interpretation of ATLAS
mono-jet analysis of Ref.~\cite{Aaboud:2016tnv} for  3.2~fb$^{-1}$
(solid blue) and projected 300~fb$^{-1}$ (dashed blue)
integrated luminosities.}
\end{figure}

In Fig.~\ref{fig:cs} we present the cross sections for the torsion mono-jet signature versus $M_{ts}$ at
LHC@13 TeV with a $\MET>100$~GeV cut, for two values of $\xi$: $\xi=0.01$ and  $\xi=\xi_{max+}$. In this
evaluation we have used the QCD renormalisation and factorisation scales, $Q$ equal to the transverse
momentum of the pair of DM particles i.e. missing transverse momentum, \MET{}, while the parton density
function,  PDF was chosen to be NNPDF23LO (\verb|as_0119_qed|) PDF set~\cite{Ball:2012cx}. In the same
Figure we present LHC  limits obtained using an interpretation of the ATLAS mono-jet analysis of
Ref.~\cite{Aaboud:2016tnv} for  3.2~fb$^{-1}$ integrated luminosity. For our analysis we performed
parton level simulation with CalcHEP, followed by PYTHIA8~\cite{Sjostrand:2007gs} and
{\sc Delphes\,3}~\cite{deFavereau:2013fsa} to simulate
hadronisation and patron showering, and for fast detector
simulation respectively. The detector level analysis was performed using {\sc CheckMATE} v2~\cite{Drees:2013wra}.
In Fig.~\ref{fig:cs} we also show the the limits on the cross section  for the projected luminosities of 300~fb$^{-1}$ for these ATLAS analysis. Our projection is based on the assumptions that the number of BG events scales with the luminosity and that the uncertainty on the BG scales as the square root of the luminosity. However, we set the lower limit for the BG uncertainty to be 1\% of the BG. This choice of 1\% for the limit on BG uncertainty is based on the post-fit numbers with respective BG error provided by ATLAS and CMS
for \MET{} bins with high statistics, see e.g.
\cite{Aaboud:2016tnv,CMS:2016tns} together with additional materials provided at
\url{http://cms-results.web.cern.ch/cms-results/public-results/preliminary-results/EXO-16-013/#AddFig}.

\begin{figure}[htb]
\centering
{\includegraphics[width=0.95\textwidth]{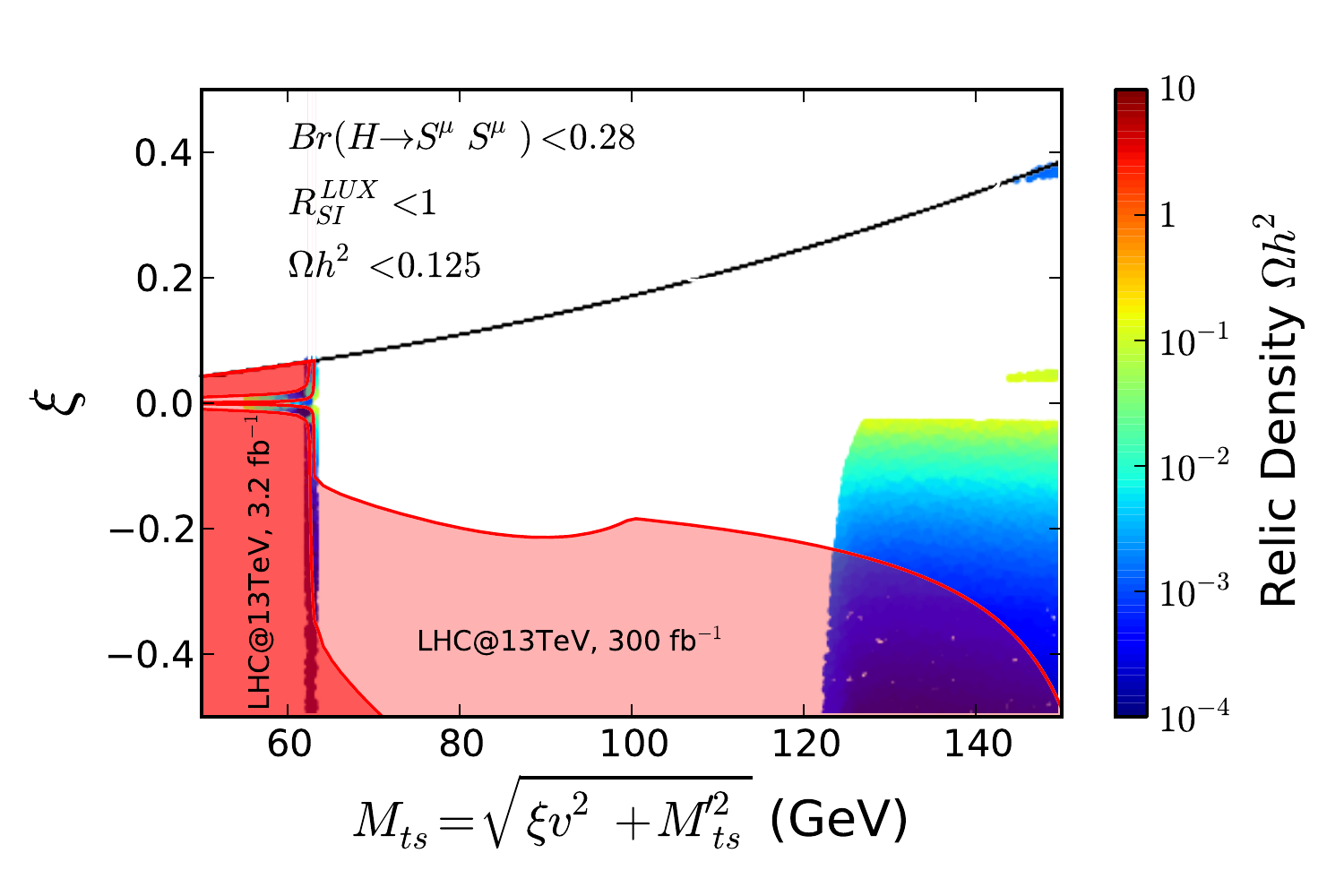}}
\vskip -0.5cm
{\includegraphics[width=0.95\textwidth]{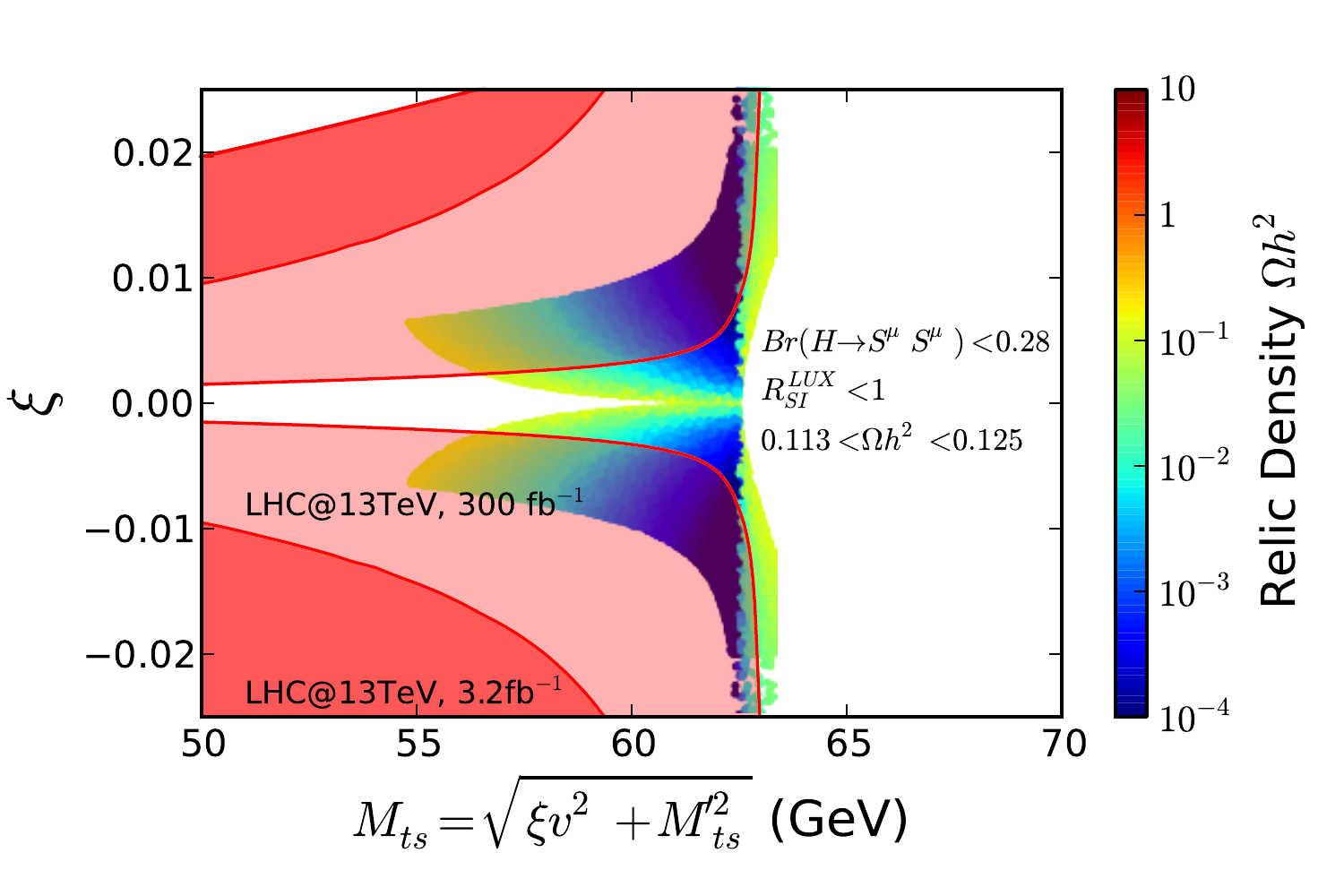}}%
\caption{\label{fig:LHC}
The LHC limits in $\xi-M_{ts}$ pane
obtained  using limits presented in  Fig.~\ref{fig:cs}.
The bottom  panel  presents the  region of $\xi-M_{ts}$ pane
zoomed  around $M_H/2$.
}
\end{figure}

From Fig.~\ref{fig:cs} one can see that even for projected 300~fb$^{-1}$ integrated luminosity
the LHC is not sensitive to the torsion parameter space for $M_{ts}>M_H/2$ and positive values of $\xi$
coupling. Moreover, data for 3.2~fb$^{-1}$ does not set any limits on $M_{ts}$
for  $\xi>0$.
In Fig.~\ref{fig:LHC} we present the LHC limits in the $\xi-M_{ts}$ plane
which we obtained  using limits presented in  Fig.~\ref{fig:cs}.
The upper panel of the figure demonstrates that the current LHC data
exclude the $M_{ts}\simeq MH/2$ parameter space only for negative values of $\xi$ below -0.35,
indicated by dark pink colour.
The bottom  panel of Fig.~\ref{fig:LHC} presents the region of the $\xi-M_{ts}$ plane
zoomed in around $M_{ts} \simeq M_H/2$ demonstrating that at 300~fb$^{-1}$
the LHC can probe
$M_{ts} < M_H/2$ masses for $|\xi|$ as low as about $2 \times 10^{-3}$, as
indicated by the light-pink colour.
From the top panel one can also see that at 300~fb$^{-1}$, the
LHC can probe the $M_{ts}>M_H$ region for $\xi\lesssim -0.2$ as well, as is 
also shown by the shaded light-pink colour.

\section{Conclusions}
\label{Sect5}

The standard argument in favour of torsion is that it provides a
geometric counterpart for the spin of matter, which looks a natural
thing in the microscopic physics. The consistency of quantum theory
of matter fields on a torsion background requires torsion to be
non-minimally coupled to both fermions of scalars of the SM.
On the other hand torsion leads to interesting particle physics
phenomenology, and also has important cosmological consequences
(see, e.g., \cite{Capozziello:2011et} for the review). We have considered a new
application which combines naturally these two aspects. The theory
of a torsion field interacting with the SM Higgs doublet and having
negligible coupling to SM fermions is protected from decaying by a
$Z_2$ symmetry and therefore becomes a promising DM candidate.

We have explored this possibility and checked the
viability of this scenario in the light of DM relic density, direct
detection and collider constraints. In the simplest  $TDM1$
scenario with just one parameter $\xi$ of the torsion-Higgs
coupling we have found that torsion can not explain 100\% of the DM
relic density because of the tension with  DM DD constraints.
However in the two parametric $\xi-M_{ts}$  $TDM2$ scenario,
when the Higgs boson is only partly responsible for generation of the
torsion mass, there is a region of the parameter space where
torsion is the sole DM. In this scenario  the mass relation
(\ref{TDM2mass}) which appears naturally within the GUT
models provides an additional degree of freedom which allows the $TDM2$
to survive DM DD constraints.

We have also found that the LHC, while not being sensitive to $TDM2$ parameter
space, anyway can further constrain it with the projected luminosity.
For example, we have demonstrated that at 300~fb$^{-1}$ the
LHC can probe masses $M_{ts}<M_H/2$ for $|\xi|$ as low as
about $2\times 10^{-3}$, including the parameter space where
torsion contributes 100\% to the DM budget of the Universe. On the
other hand at 300~fb$^{-1}$, the LHC can also probe the $M_{ts}>M_H$ scenario
for $\xi\lesssim -0.2$, although in this case torsion's contribution
to the DM relic density is at about the percent level or below.

\section*{Acknowledgements.}

The present work was started during a short-term visit of A.B. to Juiz de
Fora. The authors are very grateful to CNPq for supporting this visit.
A.B. acknowledges partial support from the STFC grant number
ST/L000296/1, NExT Institute and partial funding by a Soton-FAPESP
grant. A.B. also thanks the Royal Society Leverhulme Trust Senior
Research Fellowship LT140094.
A.B. and M.T. would like to thank FAPESP grant 2011/11973-4 for funding their visit to ICTP-SAIFR
where part of this work was completed.
I.Sh. thanks CNPq, FAPEMIG and ICTP for partial support.

\bibliography{bib}
\bibliographystyle{unsrt}
\end{document}